\documentclass[aps,prd,nofootinbib,onecolumn,superscriptaddress,showpacs]{revtex4}

\usepackage{graphicx}
\usepackage{amssymb}
\usepackage{amsmath}

\begin{document}

\title{Modifications to Lorentz invariant dispersion in relatively boosted frames}

\author{Uri Jacob}
\email{uriyada@phys.huji.ac.il}
\affiliation{Racah Institute of Physics, The Hebrew University, Jerusalem, Israel}

\author{Flavio Mercati}
\email{flavio.mercati@roma1.infn.it}
\affiliation{Dipartimento di Fisica, Universit\`a di Roma ``La Sapienza" and Sez.~Roma1 INFN, P.le A. Moro 2, 00185 Roma, Italy}

\author{Giovanni Amelino-Camelia}
\email{amelino@roma1.infn.it}
\affiliation{Dipartimento di Fisica, Universit\`a di Roma ``La Sapienza" and Sez.~Roma1 INFN, P.le A. Moro 2, 00185 Roma, Italy}

\author{Tsvi Piran}
\email{tsvi@phys.huji.ac.il}
\affiliation{Racah Institute of Physics, The Hebrew University, Jerusalem, Israel}

\begin{abstract}
We investigate the implications of energy-dependence of the speed of photons, one of the candidate effects of quantum-gravity theories that has been most studied recently, from the perspective of observations in different reference frames. We examine how a simultaneous burst of photons would be measured by two observers with a relative velocity, establishing some associated conditions for the consistency of theories. For scenarios where the Lorentz transformations remain valid these consistency conditions allow us to characterize the violations of Lorentz symmetry through an explicit description of the modification of the quantum-gravity scale in boosted frames with respect to its definition in a preferred frame.
When applied to relativistic scenarios with a deformation of Lorentz invariance that preserves the equivalence of inertial observers, we find an insightful characterization of the necessity to adopt in such frameworks non-classical features of spacetime geometry, e.g. events that are at the same spacetime point for one observer cannot be considered at the same spacetime point for other observers.
Our findings also suggest that, at least in principle (and perhaps one day even in practice), measurements of the dispersion of photons in relatively boosted frames can be particularly valuable for the purpose of testing these scenarios.
\end{abstract}


\maketitle

\section{Introduction}
Over the past few years there has been a growing interest in the investigation of possible high-energy
quantum-gravity-induced deviations from Lorentz invariance, that would induce modifications of the energy-momentum dispersion (on-shell) relation
\cite{grbgac,gampul,billetal,schaefer,kifune,mexweave,ita,aus,gactp}.
This hypothesis finds support in preliminary results obtained in some popular approaches to the study of the quantum-gravity problem, most notably approaches based on ``spacetime noncommutativity"
\cite{gacmajid,kowaNCSTdisp}
or inspired by ``loop quantum gravity"~\cite{gampul,urrutiaPRD}.
It is expected that the scale governing such deviations from Lorentz symmetry is the scale where the same frameworks predict a breakdown of the familiar description of spacetime geometry, the ``quantum-gravity energy scale", here denoted by $E_{QG}$, which should be roughly of the order of the Planck scale ($E_{pl}=\sqrt{\hbar c^5/G}\sim1.2\times10^{28}$ eV). \par

While the study of particles with energies close to the Planck scale is far beyond our reach in particle-physics laboratories and even in astrophysical observatories, it is possible to look for the minute effects that Planck-scale deviations from Lorentz symmetry produce for particles with energies much lower than the Planck scale. From this perspective of the search of small leading-order corrections, some astrophysical phenomena could provide meaningful insight, by either producing signals of the modified dispersion, or alternatively, providing constraints on the relevant models.
Indeed, astronomical observations have already set valuable limits on these leading-order corrections and as such on some of the alternative scenarios for ``quantum gravity". Most insightful are the studies based on the high-energy modifications of particle speeds \cite{grbgac,schaefer,neutNature,magicPLB,hessPRL,emnPLB2009,gacSMOLINprd,fermiNATURE}
and on the modifications to high-energy reaction thresholds (as relevant for the ultra-high-energy cosmic-rays or very-high-energy $\gamma$-rays)
\cite{kifune,ita,aus,gactp,liberatiUHECR2009,steckeNEW,absorptionPRD}.
The rapid development of astronomical instrumentation over the last years has improved significantly our ability to test these scenarios. In particular the recent results of the Fermi telescope collaboration \cite{fermiNATURE}, using a time-of-flight analysis of a short gamma-ray burst GRB090510, achieved a sensitivity to Planck-scale effects and set for the first time a limit of $E_{QG}\gtrsim E_{pl}$ (for the case of effects introduced linearly in the quantum-gravity scale, defined below as $n=1$). \par

The conceptual perspective that guides these studies is consistent with the history of other symmetries in physics, which were once thought to be fundamental but eventually turned out to be violated.
There is no essential reason to believe that Lorentz symmetry will be spared from this fate.
If the symmetry is broken at a scale $E_{QG}$, we should expect leading-order corrections to arise even at much lower energies, and hence it is just natural to explore their possible implications.
However, in view of the importance of Lorentz symmetry in the logical consistency of our present formulation of the laws of physics, one should ask which modified dispersion relations can be placed into a viable and consistent theory? The analysis presented here intends to contribute in this direction.
We investigate this question from a rather general perspective, aware of the fact that it is quantum-gravity research that provides the key motivation for these studies, but also in principle open to the possibility that the conjectured modifications of the dispersion relation might have different origin. \par

We consider a generic ``low-energy", $E\ll E_{QG}$, leading-order modification of the dispersion relation of the form:\footnote
{The factor $\frac{2}{n+1}$ is here introduced only for the consistency of $E_{QG}$ in the description of speeds with the analogous parameter most commonly used in the relevant literature.}
\begin{equation} \label{LIDform}
E^2-p^2c^2-m^2c^4 \simeq \pm \frac{2}{n+1}E^2\left(\frac{E}{E_{QG}}\right)^n .
\end{equation}
The power of this leading correction, $n$, and the parameter $E_{QG}$ are model-dependent and should be determined experimentally. The conventional speed of light constant, $c$, remains here the low-energy limit of massless particles' speed, and we put hereafter $c=1$
(with the exception of only a few formulas where we reinstate it for clarity).
Notice that the fact that we work in leading order in the Planck-scale corrections allows us to exchange the (modulus of) momentum of a photon with its energy in all Planck-scale suppressed terms.
The modification of the dispersion relation produces a difference between energy and momentum of a photon, but this is itself a first-order correction and taking it into account in terms that are already
suppressed by the smallness of the Planck scale would amount to including subleading terms. \par

The approach based on (\ref{LIDform}), which is adopted here, has been considered by many authors as the natural entry point to the phenomenology of Lorentz invariance violation \cite{grbgac,kifune,ita,aus,gactp,neutNature,magicPLB,hessPRL,liberatiUHECR2009}.
We are mainly concerned with the conceptual implications of such modifications of the dispersion relation for the way in which the same phenomenon is observed in different reference frames, and for our exploratory purposes it is sufficient to focus on (\ref{LIDform}). Our results apply to all cases in which (\ref{LIDform}) is satisfied to leading order.
The findings should apply also to scenarios with birefringence, which could be induced by Planck-scale effects (see, {\it e.g.}, Ref.~\cite{gampul}).
Our results may provide a first level of intuition even for the possibility of Planck-scale induced ``fuzziness", which is the case of scenarios in which there is no systematic modification of the dispersion relation but modifications roughly of the form (\ref{LIDform}) occur randomly, affecting different particles in different ways depending on the quantum fluctuations of spacetime that they experience (for details see, {\it e.g.}, Refs.~\cite{ngfuzzy,gacSMOLINprd}). \par

Assuming, as commonly done in the related literature~\cite{grbgac,gampul,billetal,schaefer,urrutiaPRD},
that the standard relation $v= \partial E /\partial p$ holds, the dispersion relation (\ref{LIDform}) leads to the following energy dependence of the speed of photons:
\begin{equation} \label{LawForSpeed}
v(E) = 1 \pm \left( \frac {E}{E_{QG}} \right)^n .
\end{equation}
This is an important phenomenological prediction of the modified-dispersion scenario. This effect is the basis for the most generic tests of the Lorentz-violation theories. We will consider the case of subluminal motion, corresponding to the `-' sign in Eq. (\ref{LawForSpeed}). A similar analysis can be performed for the case with superluminal speeds. We here contemplate an experiment where two relatively boosted observers detect two photons, which are emitted simultaneously at the source but arrive separated by a delay due to their different energies according to (\ref{LawForSpeed}).
The basic idea of our study is to compare, under different theoretical scenarios, the time delay measured in the different reference frames. This primarily serves as a gedanken experiment that establishes features which modified-dispersion models must include in order to have a consistent scenario. \par

We show that if in all reference frames photons have a dispersion relation as in (\ref{LIDform}) and a speed law as in (\ref{LawForSpeed}) (with fixed frame-independent parameters), then assuming the transformations between frames are governed by the standard Lorentz laws would lead to a contradiction between the measurements of the different observers. This is expected, but we use our explicit derivation to deduce other requirements the models must fulfill for a solution. We consider two classes of modified-dispersion models.
The first class consists of theories where the Lorentz Symmetry is Broken (LSB) with the existence of a preferred inertial frame (sometimes attributed to the Cosmic Microwave Background frame).
The small-scale structure and hence high-energy behavior is defined in this reference frame. Therefore, the dispersion relation is allowed to take different forms in other frames. In the following we will consider only the most common LSB theories in which the Lorentz transformations are still applicable when going from one frame to another.
The second class of models describes a scenario where the Lorentz symmetry is merely deformed, such that the equivalence of all inertial observers is maintained. This means that the Lorentz symmetry makes way for a more complex symmetry, but the theory is still relativistic (there is no preferred reference frame). These are called ``Doubly Special Relativity" (DSR) models. Within this scenario the laws of transformation are necessarily modified from the standard Lorentz transformations. This is done in such a way that all observers agree on the physical laws, including the energy-momentum dispersion relation. Not all models of DSR predict energy-dependent velocities of photons, but we deal with the analysis of the common DSR framework where the velocities behave as in (\ref{LawForSpeed}).
This is to say that we will have two complementary scenarios - LSB with standard Lorentz transformations but allowing for departures between the dispersion relations in different frames, and DSR where the theory is frame-independent but the transformations between reference frames differ from the Lorentz ones. \par

We observe that the large effort recently devoted to the phenomenology of modifications to Lorentz invariance has focused on analyses performed in a single frame. We believe that deeper insight on the fate of the Lorentz symmetry can be gained by also investigating how the same phenomenon is viewed by two different observers. We argue that our study sheds light on several conceptual issues, which in turn may well provide guidance even for the ongoing effort utilizing the standard ``laboratory frame" tools.
We obtain definitive results for our LSB scenario, and for the DSR case we find that some non-classical features of spacetime are required (while the phenomenological DSR results are severely conditioned by our restrictive assumption of a classical-spacetime implementation).
In principle (and perhaps one day in practice, after a Lorentz-violating effect is observed in one frame) our description of the effects of the LSB scenario for observations of the same phenomenon in two reference frames with a relative boost could be even exploited experimentally.
In spite of the limitations associated with the assumption of a classical-spacetime implementation for the DSR case, our findings for the comparison between the LSB scenario and the DSR scenario provide some encouragement for the possibility that our scheme could also be exploited to discriminate between these alternative spacetime-propagation models.

\section{A kinematical consistency condition}

\subsection{The general condition}
We turn now to the comparison between different observers which  is the key point in our analysis. Let us start by stressing that we shall focus for simplicity on flat (Minkowskian) spacetime. While in actual studies of astronomical signals the cosmological curvature can play a significant quantitative role in the actual time delay~\cite{CosmoDelJCAP}, we are here mainly concerned with a conceptual  analysis and curvature will not change our qualitative results. \par

In a flat spacetime it follows immediately from (\ref{LawForSpeed}) that two photons with energies $E_1$ and $E_2$, emitted simultaneously by a distant source, will not reach the observer simultaneously.
For a source at distance $c T$ from the observer, one should find a difference in arrival times at the detector given by:
\begin{equation} \label{TimeDelayLinear}
\Delta T = T ~\left( \frac {E_2^n - E_1^n}{E_{QG}^n} \right) .
\end{equation}
Clearly then, if the time of arrival of a low-energy photon is $T$, the time of arrival of a photon with high energy $E$ will be $T_E = T + T \left( \frac {E}{E_{QG}} \right)^n$. \par

We are interested in the comparison between the measurements of the same photons, emitted from a distant source, by two observers, $\mathcal{O}$ and $\mathcal{O}'$, in relative motion. We consider a photon of low energy $E_1$ and a photon of high energy $E_2$. In addition to the mentioned assumption of flat spacetime, we assume that the two photons are emitted simultaneously from a single spacetime point. For simplicity we consider a source, $\mathcal{S}$, at rest\footnote
{In fact the source's motion is irrelevant for the analysis, and all that matters is that the two photons are emitted at a single spacetime point for any observer. This assumption is valid in classical spacetimes (see discussion later).}
with respect to the observer $\mathcal{O}$. We denote by $d$ the spatial distance from the source as measured by $\mathcal{O}$ ($d = c T$). We denote by $x_s$ and $x_d$ the position coordinates of the source and the detector respectively, and we denote by $\beta$ the relative speed between observers $\mathcal{O}$ and $\mathcal{O}'$. The two observers are at the same spacetime point, where the faster photon reaches both of them, so they are synchronized there, {\it i.e.} $x_d=x_d'=0$ at $t_1=t_1'=0$.
We also implicitly assume that the synchronization of clocks between the two observers is essentially standard: they would perform clock synchronization using low-energy photons\footnote
{In principle one has to consider also the inherent uncertainty associated with the time measurement of low-energy photons, but this can be neglected when we consider time delays induced by distant sources.},
which are practically unaffected by the modification of the dispersion relation. \par

\begin{figure}[ht]
 \includegraphics[width=\textwidth]{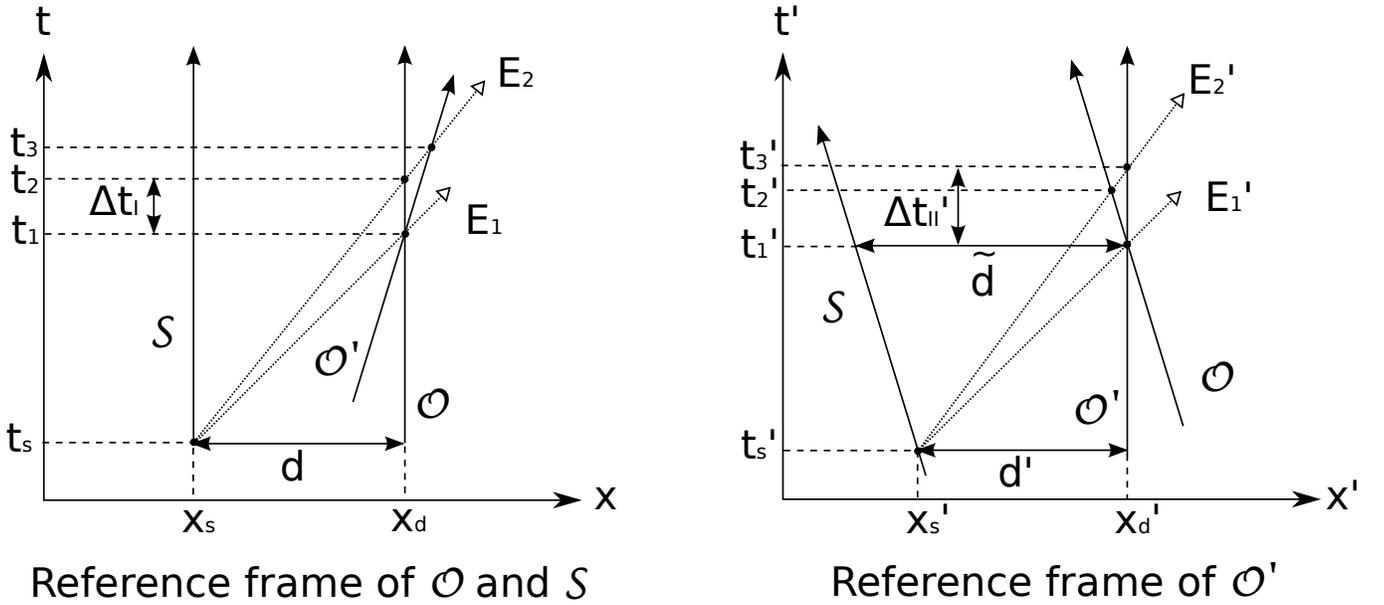}
 \caption{Space-time diagrams of the dispersion of photons in two reference frames.}
 \label{figure1}
\end{figure}
\begin{figure}[ht]
 \includegraphics[width=0.5\textwidth]{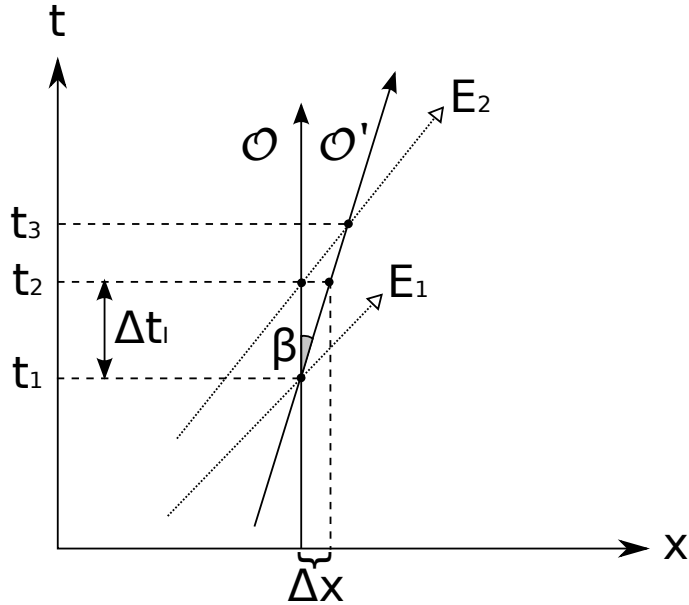}
 \caption{Zoom of the space-time diagram in the reference frame of $\mathcal{O}$, with $\Delta x$ highlighted.}
 \label{figure2}
\end{figure}

In Fig. \ref{figure1} one can see the space-time diagrams in the
two reference frames, the first at rest with respect to $\mathcal{O}/\mathcal{S}$
and the second at rest with respect to $\mathcal{O'}$. Only the case in which $\mathcal{O}'$ is
moving away from $\mathcal{S}$ is depicted, but in the formulas a ``$\pm$'' in front of
$\beta$ will clarify, where needed, that the formulas are valid also in the case in which
$\mathcal{O}'$ is moving towards $\mathcal{S}$.
Each observer measures the delay in the time of arrival of the photon $E_2$
with respect to the photon $E_1$ in its own reference system. This means that $\mathcal{O}$ measures $\Delta t_I = t_2-t_1$, while $\mathcal{O}'$ measures
$\Delta t_{II}' = t_3' - t_1'$. The measurement of $\mathcal{O}'$ appears dilated
in the reference frame of $\mathcal{O}$, where it is $\Delta t_{II} = t_3 - t_1$.
$t_3-t_2$ is the time needed by the photon moving at speed $v(E_2)$ to catch up with the
observer $\mathcal{O}'$, that moves at speed $\beta$ and has an advantage of
$\Delta x = \pm \beta ~ (t_2 - t_1)$ (this last relation can be immediately verified
by looking at Fig. \ref{figure2}).
 \par

The kinematics in the reference frame $\mathcal{O}$ yields the relation:
\begin{equation}
t_3 = t_2 + \frac{\Delta x }{v(E_2) \mp \beta} = t_2 \pm \frac{\beta }{v(E_2) \mp \beta} (t_2-t_1) .
\end{equation}
Subtracting both sides by $t_1$ and recalling that $\Delta t_{I} = t_2 - t_1$ and
$\Delta t_{II} = t_3 - t_1$, we get (the ``$-$'' sign refers to the case of $\mathcal{O}'$
moving away from $\mathcal{S}$):
\begin{equation} \label{ConsCond}
\Delta t_{II} = \Delta t_{I} \left( \frac{v(E_2)}{v(E_2) \mp \beta} \right) .
\end{equation}
This relation is central to our analysis, since it was derived in the reference frame $\mathcal{O}$ (making no assumptions regarding the behavior of other reference frames). The right-hand-side of (\ref{ConsCond}) can be expressed using exclusively the law of energy dependence of the speed of photons in the first rest frame (of $\mathcal{O}$ and $\mathcal{S}$). The formula can be used to relate the results of time-delay measurements performed by the two observers $\mathcal{O}$ and $\mathcal{O}'$, respectively $\Delta t_{I}$ and $\Delta t_{II}'$. This will impose limits on the free parameters, the speed law in the reference frame $\mathcal{O}'$ and the laws of transformation between observers. \par

Regarding the boost transformation that acting on $\Delta t_{II}$ gives $\Delta t_{II}'$, we can write:
\begin{equation} \label{ConsCondBOOST}
{\cal N}_\beta^{-1} [\Delta t_{II}']
= \Delta t_{I} \left( \frac{v(E_2)}{v(E_2) \mp \beta} \right) ,
\end{equation}
where we denoted formally by ${\cal N}_\beta^{-1}$ the action of the (inverse) boost that takes from $\mathcal{O}'$ to $\mathcal{O}$. ${\cal N}_\beta$ reduces to the Lorentz transformation in the case of Lorentz invariance, and if the photons' speed is independent of energy this condition is trivially satisfied. ${\cal N}_\beta$ is also the usual Lorentz transformation in our analysis of LSB. However, it takes a more general form in the case of DSR.
The relation (\ref{ConsCondBOOST}) should be viewed as a requirement of logical consistency:
if all the assumptions we have made hold and there exist modified dispersion laws and a transformation action ${\cal N}_\beta$,\footnote
{We discuss the possibility of a quantum spacetime, where some assumptions we have made which imply the classicality of spacetime cannot be adopted, later in the text.}
then inevitably they must be connected through (\ref{ConsCondBOOST}). \par

\subsection{Analysis with classical boost transformations} \label{ClasAnalSec}
If the speed of photons depends on their energy, but we keep the classical transformation laws between observers unmodified, it is easy to see that the law giving the quantitative dependence of the speed of photons on energy must be observer-dependent.
We demonstrate explicitly in this subsection that if ${\cal N}_\beta$ is the standard Lorentz transformation action and both $\Delta t_{I}$ and $\Delta t_{II}'$ obeyed the same frame-independent but energy-dependent speed law, the consistency condition would not be met.
This derivation with Lorentz transformations will then be used in Sec. \ref{LSBsec} as the basis for the LSB analysis. \par

If the transformation laws between the reference frame of $\mathcal{O}$
and that of $\mathcal{O}'$ are the special-relativistic ones, since the delay
$\Delta t_{II}'$ is a proper time in the reference frame of $\mathcal{O}'$, it
is related to $\Delta t_{II}$ by the usual time-dilation formula:
\begin{equation}
\Delta t_{II} = {\cal N}_\beta^{-1} [\Delta t_{II}'] =
\gamma \Delta t_{II}'
\end{equation}
(where as usual $\gamma \equiv 1/\sqrt{1-\beta^2}$).
We use this to rewrite Eq. (\ref{ConsCond}) as follows:
\begin{equation} \label{jocz}
\Delta t_{II}' = \Delta t_{I} \left( \frac{v(E_2)}{v(E_2) \mp \beta} \right) \gamma^{-1} .
\end{equation}
Next we notice that in this equation $\Delta t_{I}$ is of order $E_{QG}^{-n}$
(clearly it must be the case that $\Delta t_{I} \rightarrow 0$ if $E_{QG} \rightarrow \infty$).
Therefore, for an analysis in leading order of $E_{QG}^{-n}$,
where we also assume $1-v(E_2) \ll 1-\beta$,
it is possible to take $v(E_2) \approx 1$ on the right-hand-side of (\ref{jocz}):
\begin{equation} \label{DeltaT2PrimoBroken}
\Delta t_{II}' = \Delta t_{I} \left( \frac{\gamma^{-1}}{1\mp \beta} \right)
= \mathcal{D}^{-1} \Delta t_{I} ,
\end{equation}
where we denoted by
\begin{equation}
\mathcal{D} = \sqrt{\frac{1 \mp \beta}{1 \pm \beta}} ,
\end{equation}
the relativistic Doppler factor. \par

Equation (\ref{DeltaT2PrimoBroken}) gives the time delay $\Delta t_{II}'$ that $\mathcal{O}'$
must observe for consistency with the time delay $\Delta t_I$ seen by $\mathcal{O}$.
We should describe $\Delta t_{II}'$ in terms of quantities measured
by $\mathcal{O}'$ itself, so that we can investigate the form of the time-delay formula
according to $\mathcal{O}'$ on the basis of the form postulated for $\mathcal{O}$.
First, using the speed law of (\ref{LawForSpeed}) in the reference frame of $\mathcal{O}$, as explained, we explicitate $\Delta t_I$ in terms of the energies and the distance measured by $\mathcal{O}$:
\begin{equation} \label{DelO}
\Delta t_{I} = d ~\frac{E_2^n - E_1^n}{E_{QG}^n} .
\end{equation}
Putting this in (\ref{DeltaT2PrimoBroken}) we have:
\begin{equation} \label{jocw}
\Delta t_{II}' = \mathcal{D}^{-1} ~d ~\frac{E_2^n - E_1^n}{E_{QG}^n} .
\end{equation}
Now we can use the fact that
the relativistic Doppler factor also connects the energy $E$ of a particle as seen by
observer $\mathcal{O}$ and the corresponding energy $E'$ seen by observer $\mathcal{O}'$,
$E'= \mathcal{D} E$, so that:
\begin{equation} \label{jocu}
E_2^n - E_1^n = \left[ E_2'^n - E_1'^n \right] \mathcal{D}^{-n} .
\end{equation}
Since $d$ is already multiplied by a factor $E_{QG}^{-n}$ in (\ref{jocw}), we can, consistently to leading-order, ignore all effects of $E_{QG}$ on the relationship between $d'$ and $d$.
Accordingly, we can handle $d$ as the space component of a light-cone vector (that is the vector connecting the point with space-time coordinates $\{t_s,x_s\}$ to the point with coordinates $\{t_1,x_d\}$, see Fig. \ref{figure1}), and so it also transforms through the Doppler factor:
\begin{equation} \label{jocv}
d' = \mathcal{D} d .
\end{equation}
The same conclusion is of course drawn if we observe that
according to $\mathcal{O'}$ the distance between $\mathcal{S}$ and
$\mathcal{O}$ is $\tilde{d} = d/\gamma$ (standard Lorentz contraction of the proper length $d$), but in order to obtain $d'$ (the distance traveled by the photon according to $\mathcal{O'}$ - the distance between $\mathcal{O'}$ and $\mathcal{S}$ at emission) one must consider also the distance that according to $\mathcal{O'}$ the source $\mathcal{S}$ travels during the time $d'/c$:
\begin{equation}
\tilde{d} = d' \pm \beta d' \Rightarrow
d' = \frac{\tilde{d}}{1 \pm \beta} = \frac{d}{\gamma (1 \pm \beta)} = \mathcal{D} d ,
\end{equation}
in agreement with (\ref{jocv}). \par

Substituting (\ref{jocu}) and (\ref{jocv}) in (\ref{jocw}), we can finally describe the
time delay seen by observer $\mathcal{O'}$ in terms of other quantities
measured by $\mathcal{O'}$:
\begin{equation} \label{sec2final}
\Delta t_{II}' = \mathcal{D}^{-n-2} \left( d' \frac{E_2'^n - E_1'^n}{E_{QG}^n} \right) .
\end{equation}
Comparing (\ref{DelO}) with (\ref{sec2final}), we see that in such a scenario the speed law is not frame-independent. The modified dispersion relation is inconsistent with keeping the two familiar ingredients of frame independence and Lorentz transformation laws.
We have to abandon at least one of these appealing notions in order to have a consistent solution. The following sections discuss two different scenarios that can achieve consistency by allowing for these additional departures from special-relativity: (i) breaking the symmetry and having a ``preferred-frame scenario" (Sec. \ref{LSBsec}) or (ii) modifying the Lorentz transformations (Sec. \ref{DSRsec}).

\section{The broken-symmetry scenario} \label{LSBsec}
We have found that if we use the standard Lorentz transformations, we necessarily reach a LSB ``preferred-frame scenario". This means that in principle the theory could be formulated in one chosen frame of reference (typically one where the laws take an appealingly simple form) and then all other observers see the ``Lorentz-transformed image" of the fundamental laws written in the preferred frame.
This is indeed the conceptual perspective adopted in the most studied Lorentz-invariance-violation models, where the scale $E_{QG}$ in (\ref{LawForSpeed}) is expected to take different values for different observers.
It is valuable to establish how exactly the laws change form in going from one reference frame to another, since this is the only way to combine the results of experiments performed by different observers.
The result of the previous derivation turned out to provide a quantitative characterization of the frame dependence of the LSB scenario. We shall here continue that derivation, aware that it must be viewed in the context of an observer-dependent framework, and obtain further characterizations of the consistent broken-symmetry scenario. \par

According to the analysis of Sec. \ref{ClasAnalSec}, if we assume that the Lorentz transformations hold and in a certain (preferred) frame the speed of photons behaves as in (\ref{LawForSpeed}), the photon speed law must be quantitatively different for relatively boosted observers.
The result (\ref{sec2final}), derived on the basis of our logical-consistency requirement, implies that
starting from the speed law $v(E) = 1-\left( \frac{E}{E_{QG}} \right)^n$ in a frame $\mathcal{O}$, one arrives at the following speed law in a frame $\mathcal{O'}$:
\begin{equation} \label{LawForSpeedPrimeBroken}
v'(E') = 1 -\mathcal{D}^{-n-2} \left( \frac{E'}{E_{QG}} \right)^n .
\end{equation}
We therefore have an explicit characterization of the expected change of the in-vacuo-dispersion
between reference frames.
The laws in $\mathcal{O}$ and in $\mathcal{O}'$ have the same functional form, but they differ quantitatively with the addition of the Doppler factor parameter in $\mathcal{O}'$. This parameter is determined in each frame by its relative speed with respect to the privileged frame - for any observer it is fixed and the addition is just a numerical constant. This allows us to treat the difference between observers as a variation in the scale of symmetry-breaking - marking with $E_{QG}$ the scale  as measured by $\mathcal{O}$, one then finds that the scale as measured by $\mathcal{O}'$ is:
\begin{equation} \label{jocccc}
E_{QG}' = \mathcal{D}^{1+2/n} E_{QG} .
\end{equation}
This definitive description of the quantum-gravity-scale transformation is in itself an important characterization. It is especially noteworthy as several authors had expected (see, {\it e.g.}, Ref.~\cite{urrucontraction} and references therein) that $1/E_{QG}$ should be treated as a length scale, subject to standard Lorentz-FitzGerald contraction. Equation (\ref{jocccc}) shows that this is not the case. \par

Another potentially valuable characterization is found upon recalling the assumption that the energy dependence of the speed of photons, (\ref{LawForSpeed}), is describable as the group velocity of photons governed by a deformed dispersion relation:
\begin{equation}
E^2-p^2 = -\frac{2}{n+1} E^2 \left( \frac{E}{E_{QG}} \right)^n .
\end{equation}
We now deduce from Eq. (\ref{LawForSpeedPrimeBroken}) with $v'(E')= \partial E' /\partial p'$,
that in the reference system of $\mathcal{O}'$ the dispersion relation takes the form:
\begin{equation}
E'^2 - p'^2 = - \frac{2}{n+1} E'^2 \left( \frac{E'}{E_{QG}} \right)^n \mathcal{D}^{-n-2} = -\frac{2}{n+1} (E'/\mathcal{D})^2 \left( \frac{E' /\mathcal{D}}{E_{QG}} \right)^n =
-\frac{2}{n+1} E^2 \left( \frac{E}{E_{QG}} \right)^n .
\end{equation}
Here on the right-hand-side we have highlighted a peculiar consequence of the law of transformation of the scale $E_{QG}$. Its net result is that, for a given photon (of given energy $E$ according to observer $\mathcal{O}$), the violation of the special-relativistic dispersion relation, $E^2-p^2=0$, has exactly the same magnitude for all observers. This magnitude is fixed by the formula for the Lorentz-violating term in the preferred frame, which is based on the photon energy $E$, measured in that frame:
\begin{equation}
\mu^2 \equiv - \frac{2}{n+1} E^2 \left( \frac{E}{E_{QG}} \right)^n .
\end{equation}
Any observer can compute the Lorentz-violating term of the dispersion relation by finding the ``special" energy $E$, according to its boost relative to $\mathcal{O}$, and using the same formula. \par

We have evidently found, in the context of our frame-dependent scenario, an invariant term, which has the form of a mass  term in special-relativity. We stress that obviously this invariant $\mu^2$
cannot be viewed as an ``effective mass for photons in the LSB scenario", as already implied by the fact that the speed law (\ref{LawForSpeedPrimeBroken}) is not obtained by treating $\mu$ as a mass in the ultrarelativistic limit:\footnote
{An insightful way to characterize our $\mu^2$ is found by observing that $\mu^2$ is an invariant under passive Lorentz transformations but not an invariant under active Lorentz transformations, something that is not at all surprising for cases where Lorentz symmetry is broken. Under passive boosts that establish how the properties of the same particle are measured by different observers, we have indeed found that $\mu^2$ is an invariant.
Of course the observation reported in (\ref{jocjoc}) simply reflects the evident fact that $\mu^2$ is not invariant under active boosts, which for a single observer map a particle of energy $E_1$ into a different particle of energy $E_2$.
Another example of corrections that are invariant under passive Lorentz transformations but not invariant under active Lorentz transformations is the case of correction terms of the form $W_\mu p^\mu$, with $W_\mu$ some ``external" Lorentz-symmetry-breaking vector and $p_\mu$ the four-momentum of the particles: under a passive boost, of course, both $W_\mu$ and $p_\mu$ would transform in a way just such $W_\mu p^\mu$ is left invariant, but under an active boost only $p_\mu$ is transformed and the term $W_\mu p^\mu$ changes its value.}
\begin{equation} \label{jocjoc}
v'(E') \neq 1 - \frac{\mu^2}{2E'^2} .
\end{equation}
Of course, a generic correction $\mu^2$ to the dispersion relation will affect inertia in the way prescribed for mass only if, unlike our $\mu^2$, it is independent of energy-momentum (so that it would provide inertia in standard fashion in the group-velocity calculation, obtained by $\partial E/\partial p$). One similarity between our case and the analogy of a massive-particle dispersion is that in transforming to a reference frame where the energy of the particle is smaller, the effect of the correction to $v(E)=1$ becomes larger. This result for the case of LSB comes in contrast to the intuition, by which the phenomenological effect of Lorentz invariance violation should become more significant when a particle has a larger energy which is closer to the quantum-gravity scale. \par

In conclusion, regarding the much-studied quantum-gravity scenario of a high-energy breakdown of Lorentz symmetry with the emergence of a preferred frame, we have formulated explicitly the only allowed form of the modified dispersion relation, which is consistent with the Lorentz transformations.
To have a quantitative appreciation of the phenomenological effect of the Lorentz-violating speed law in different reference frames, we observe again the result, (\ref{DeltaT2PrimoBroken}), obtained for the time delay that would be measured in a general frame. For observations made from different reference frames with a relative speed $\beta$, the time delays between the photons will differ by a factor of the corresponding $\mathcal{D}$. Since the difference between the reference frames is not suppressed by another power of $E/E_{QG}$ (but only by the smallness of the relative speed), it may in principle be an observed phenomenon. As an example, we may consider a detector moving at a speed of $70$ km/s (the speed of NASA's Helios 2 satellite, which was launched in the 1970s): $v=70km/s=2.3\times10^{-4}c \Rightarrow \mathcal{D}^{-1}=1.00023$ (when the detector is moving away from the photons). This means that if, for instance, a detector at ``rest" measures a Lorentz-violation-induced time delay of $1$ s (which the Fermi telescope can measure if indeed a $n=1$ effect exists), the moving detector will measure a time delay greater by $0.23$ ms. This is small but not unrealistic to expect a possible test in the future. Note that our result is that in the LSB context the time delay in the moving frame is larger even though the photons' measured energies are lower. This is in contrast to the intuitive expectation of a smaller time delay in a frame where the photon energy is smaller (see Sec.~\ref{ClosingSec} for further discussion).

\section{The Deformed-symmetry scenario} \label{DSRsec}

\subsection{Preliminaries}
A different scenario, which has recently generated interest - the DSR scenario, was essentially introduced~\cite{gacIJMPD11,gacPLB2001} to explore the possibility that the quantum-gravity scale may affect the laws of transformation between observers while preserving the relativistic nature of the theory (no preferred frame).
Within this DSR scenario the departures from ordinary Lorentz invariance would take the shape of a ``deformation" (rather than breakdown) of symmetries. This can be thought of in a close analogy with the transition from Galilean invariance to Lorentz invariance. Galilean invariance
ensures the equivalence of inertial observers but makes no room
for observer-independent relativistically-non-trivial scales.
Attempts to accommodate a maximum-speed law first relied on
scenarios that would break Galilean symmetry (ether),
but eventually found proper formulation in the shape
of a deformation of Galilean invariance. This formulation is Lorentz invariance,
which is still without a preferred frame, but the laws of transformation between
inertial observers are deformed (with respect to the Galilean case)
in such a way that a velocity scale, setting the speed-of-light maximum value, is observer-independent.
The DSR framework makes another step with a deformation of Lorentz invariance that introduces an
additional observer-independent scale. The aim here is exploring the possibility that the existence
of a short-distance (high-energy) relativistically-invariant scale might eventually become manifest
as we probe high-energy regimes more sensitively, just like the presence of a relativistic velocity
scale becomes manifest in accurate observations of high-velocity particles. \par

We are interested in considering the case of DSR models in which the new observer-independent
law is a law of in-vacuo dispersion, based on a relativistic energy scale. This will offer us
a framework for investigation which is complementary to our LSB case. In the LSB scenario one
assumes the laws of transformation to remain unchanged, allowing for an observer-dependent
scale of in-vacuo dispersion, while in a DSR scenario with in-vacuo dispersion one must insist
on observer independence of the scale of in-vacuo dispersion, allowing the laws of
transformation between observers to take a new form. \par

Our analysis is somewhat limited by the fact that these DSR scenarios are still ``work in progress". Among the residual grey areas for a DSR description of in-vacuo dispersion particularly significant for our purposes are the ones concerning the description of spacetime.
In the following we assume a classical-spacetime geometry, within which we adopt, as done by most authors in the field~\cite{gacmajid,kowaNCSTdisp,rainbowDSR}, concepts such as spacetime points, translational symmetries, energy, momentum and group velocity in the familiar manner. We find, below, that in-vacuo dispersion of the form (\ref{LIDform}) is inconsistent with DSR in such a classical spacetime. This provides new and particularly strong elements in favor of a non-classical spacetime picture.
These findings are consistent with several independent arguments giving indirect evidence of the necessity to adopt a novel, ``quantum", picture of spacetime (see, {\it e.g.}, Refs.~\cite{aurelio,kowaNCSTdisp,rainbowDSR}). In such a quantum spacetime the sharp absolute concept of spacetime would be lost, and in particular one might have that two events that are simultaneous at the same spatial point for one observer are not simultaneous for another observer. However, a satisfactory picture of such a quantum spacetime has not yet been found. \par

The DSR side of our analysis also intriguingly suggests that quantitatively important differences can be found between the LSB and DSR cases when a single phenomenon is observed from two different reference frames. While clearly conditioned by our restrictive assumptions, and therefore subject to further scrutiny in future DSR studies, the magnitude of the effect is large enough that it would be surprising if more refined analyses removed it completely. We shall argue that this might open the way to a future test for the discrimination between alternative in-vacuo dispersion scenarios (provided that in-vacuo dispersion is detected). \par

\subsection{Analysis with observer-independent in-vacuo dispersion}
We proceed, inspecting the possibility of the DSR scenario with in-vacuo dispersion, using the logical-consistency criterion derived earlier.
We here consider the difference in time delays seen by two observers as a result of in-vacuo dispersion in the DSR case. Of course, as long as we consider a single observer there is no distinction between LSB-case and DSR-case in-vacuo dispersion, and also in the DSR scenario we have for observer $\mathcal{O}$:
\begin{equation} \label{DeltaT1joc}
\Delta t_{I} = d ~\frac{E_2^n - E_1^n}{E_{QG}^n} .
\end{equation}
The differences between the LSB and DSR scenarios are instead very significant
for what concerns the relationship between this formula for observer $\mathcal{O}$
and the corresponding formula for an observer $\mathcal{O}'$ boosted with respect to $\mathcal{O}$.
We now start from the axiom of our DSR scenario, that $E_{QG}$ is an observer-independent scale and the speed law of photons is identical in all reference frames. This is in contrast to the property of the LSB case that enabled a consistent scenario. Here, we seek a consistent description through a new form of transformation laws between observers.\footnote
{Generally in the DSR framework, we have non-standard transformations which can take a range of complex forms~\cite{gacIJMPD11,gacPLB2001}. The DSR models outline the transformation laws in energy-momentum space, but while there exist  definite transformations of energy-momentum, there is no explicit characterization of the transformations of space-time coordinates. Assuming that such transformations exist in the classical sense, then in zeroth order (for $E \ll E_{QG}$) the transformations should converge to the standard Lorentz transformations, and in leading order there would be some correction term with $E_{QG}^{-n}$.}
Exploiting the observer independence of the speed law (\ref{LawForSpeed}), we directly arrive at the time delay measured by observer $\mathcal{O}'$:
\begin{equation} \label{DeltaT2PrimoPreservedSHORT}
\Delta t_{II}' = d' \frac{E_2'^n - E_1'^n}{ E_{QG}^n} .
\end{equation} \par

The measurements in the different reference frames, (\ref{DeltaT1joc}) and (\ref{DeltaT2PrimoPreservedSHORT}), have to be related through the criterion (\ref{ConsCondBOOST}) for consistency. This clearly requires a departure from the standard Lorentz transformations. For the purpose of comparing $\Delta t_{I}$ and $\Delta t_{II}'$ we should re-express the formula (\ref{DeltaT2PrimoPreservedSHORT}) in terms of energies and distances measured by observer $\mathcal{O}$. We should, generally, write the relations between $d$ and $d'$ and between $E$ and $E'$ with a correction to the Lorentz transformations. However, $\Delta t_{II}'$ already contains a factor $E_{QG}^{-n}$, so within our leading-order analysis the zeroth order expressions will suffice, {\it i.e.} we can use
\begin{equation}
d' = \mathcal{D} d + \mathcal{O}\left(E_{QG}^{-n}\right) ~,~~
E_2'^n - E_1'^n = \mathcal{D}^{n} \left[ E_2^n - E_1^n \right] + \mathcal{O}\left(E_{QG}^{-n}\right) ,
\end{equation}
which yields:\footnote
{Note that this formula agrees, for $n=1$, with the suggestion of Ref.~\cite{hossebox} that was based on a qualitative argument.}
\begin{equation} \label{DeltaT2PrimoPreserved}
\Delta t_{II}' = \mathcal{D}^{n+1} d \frac {E_2^n-E_1^n}{E_{QG}^n} + \mathcal{O}\left(E_{QG}^{-2n}\right) = \mathcal{D}^{n+1} \Delta t_I + \mathcal{O}\left(E_{QG}^{-2n}\right) .
\end{equation}
Utilizing now the consistency criterion,
\begin{equation} \label{jocDELTADELTA}
\Delta t_{II} = \Delta t_{I} \left( \frac{1}{1 \mp \beta} \right) ,
\end{equation}
we arrive at a formulation of the connection between the time interval $\Delta t_{II}$ and its
transformation to the frame of observer $\mathcal{O}'$:
\begin{equation} \label{DSRtransLawDeltat2}
\Delta t_{II}' = {\cal N}_\beta [\Delta t_{II}] = \mathcal{D}^{n+1} (1 \mp \beta) \Delta t_{II} + \mathcal{O}\left(E_{QG}^{-2n}\right) .
\end{equation} \par

This last result poses an immediate challenge for the consistency of this scenario.
The corresponding formula for the Lorentz transformation of the proper time $\Delta t_{II}'$ is:
\begin{equation} \label{srREF}
\Delta t'_{II}\big|_{Lorentz-transfomation} = \gamma^{-1} \Delta t_{II} .
\end{equation}
It is obvious that the expression (\ref{DSRtransLawDeltat2})
cannot be a leading-order correction of (\ref{srREF}). The
crucial observation is that the difference between the
transformation laws of (\ref{DSRtransLawDeltat2}) and
(\ref{srREF}) is of zeroth order, {\it i.e.} it is not proportional
to any powers of $E_{QG}^{-1}$.
In our analysis we tried to meet the consistency requirement by allowing for some
deformations of the transformation laws, such that at zeroth order the DSR
transformations are the same as the special-relativistic ones. However, higher-order corrections
of the transformations were not relevant in our derivation, which consisted only of quantities
which are already suppressed by $E_{QG}^{-n}$. Since the only essential assumption that
was used to derive this paradoxical result concerned the classicality of spacetime geometry,
we conclude that there cannot be any consistent classical-spacetime formulation
of DSR with modified dispersion. \par

\subsection{Interpretation}
The comparison of (\ref{DSRtransLawDeltat2}) and (\ref{srREF}) allowed us to conclude, consistently with indications that emerged in previous DSR studies\footnote
{We should stress a difference between
the studies reported in \cite{rainbowDSR,dsrIJMPrev,aurelio}
and in \cite{unruhDSR,hossebox}: all of these studies expose inadequacies
of a classical-spacetime picture for DSR schemes with energy dependence of the speed of
photons, but \cite{unruhDSR,hossebox} essentially adopt as working assumption
that any novel geometric description of spacetime could at best affect
the structure of a spacetime point only quasi-locally (with fuzziness confined to a relatively small neighborhood
of the approximately associated classical-spacetime point),
whereas \cite{rainbowDSR,dsrIJMPrev,aurelio} admit more general possibilities
for spacetime quantization. A valuable formalization of this latter perspective
can be found in the recent Ref.~\cite{leeTWOTIMES}.}
\cite{rainbowDSR,dsrIJMPrev,aurelio,unruhDSR,hossebox},
but in our opinion more forcefully than in any previous related
investigation, that a classical-spacetime formulation of DSR
models with energy-dependent photon velocity can be excluded.
We must either adopt a DSR framework with no energy-dependent photon dispersion or alternatively consider non-classical features of the DSR spacetime, such that there is no classical-algebraic-description of the transformations of spacetime coordinates between reference frames. Instead the transition to DSR should be one that involves a change to the nature of spacetime. \par

It is, indeed, not very surprising to find a contradiction between the implications of the DSR framework and the properties of classical spacetime, since the quantum-gravity motivation for DSR research
is strongly related to the concept of ``quantum spacetime".
A quantum description of spacetime entails dramatic modifications to common features of spacetime geometry that are intuitive in our classical description. The sharp observer-independent identification of an event (a spacetime point) is not available in a quantum spacetime. Instead we have some ``fuzziness". The new geometry will have a fundamentally different structure.
For example, a popular quantum-gravity-inspired description of these
spacetimes relies on the formalism of ``spacetime noncommutativity", where
the coordinates of an event are described in terms of a set of noncommuting
observables, governed for example by noncommutativity of the
type $[x_\mu , x_\nu] = i {\cal C}^\alpha_{\mu \nu} \frac{x_\alpha}{E_{QG}}$,
with ${\cal C}^\alpha_{\mu \nu}$ a model-dependent dimensionless matrix.
Among other approaches that resist the temptation of characterizing spacetime nonclassicality
in terms of coordinates, one of the most popular is
the ``rainbow metric" picture proposed in \cite{rainbowDSR},
which essentially characterizes the nonclassicality of spacetime in terms
of a metric tensor which is perceived differently by particles of different energy. \par

This formalization attempts to also find support in some results obtained
through direct analysis
of DSR scenarios with in-vacuo dispersion.
In particular, it is known~\cite{dsrIJMPrev} that
observer-independent modifications of the dispersion relation
for massive particles
are inconsistent with a classical-spacetime picture.
This is seen considering the case of two particles
of masses $m_A$ and $m_B$ with the same velocity according to some observer $\mathcal{O}$.
The observer $\mathcal{O}$ could see two particles with different masses and different energies moving at the same speed and following the same trajectory (for $\mathcal{O}$ particles $A$ and $B$ are ``near"
at all times), but, taking into account the DSR-deformed laws of transformation
between observers \cite{gacIJMPD11,dsrIJMPrev}, the same two particles would have different velocities according to a second observer $\mathcal{O}'$ (so according to $\mathcal{O}'$ they could
be ``near" only for a limited amount of time).
This would clearly be a manifestation of spacetime nonclassicality, since it amounts
to stating that a single spacetime point for $\mathcal{O}$ is mapped into two (possibly sizeably distant)
points for $\mathcal{O}'$. \par

The results so far obtained in this section
provide a novel characterization of the ``virulence" of the modification of spacetime
structure required by DSR scenarios with in-vacuo dispersion.
Our findings confirm previous indications in support of the idea
that these scenarios require a {\it soft deformation}
of energy-momentum space but a rather drastic change in the description of spacetime.
With respect to previous studies that provided support for similar characterizations,
our analysis contributes a perspective centered on time measurements,
whereas earlier works~\cite{rainbowDSR,dsrIJMPrev}
had focused mainly on spatial-position measurements.
Accordingly, our findings, particularly the form of Eq. (\ref{DSRtransLawDeltat2}),
could provide motivation for further investigation of the
definition of time in DSR scenarios.
Already in some of the first DSR studies~\cite{gacPLB2001} it was realized that
several subtle issues may affect such a definition of time,
but these challenges have not been pursued further. \par

Regarding the peculiar properties of quantum spacetime, of particular relevance for our analysis is the concept of simultaneity. Special relativity removes the abstraction of an ``absolute time", but still affords us an objective, observer-independent, concept of simultaneity for events occurring at the same spatial point.
In a quantum-spacetime setting, however, events which are simultaneous and at the same spatial point for a certain observer are not necessarily simultaneous for another observer~\cite{gacPLB2001,ngtime}.
This is easily seen by noticing that most models of spacetime quantization are inspired by the quantization of (position-velocity) phase space in ordinary quantum mechanics, where the Heisenberg principle obstructs the possibility of two particles objectively having the same ``position" in phase space (they cannot have sharp values of both position and velocity, so they cannot even be attributed sharply to a point of phase space in a classical sense).
Indeed, the related concept of a ``Generalized Uncertainty Principle", introducing an absolute limitation to combined sharp determinations of both position and time of an event, has been introduced in most approaches to the spacetime quantization~\cite{garay}. \par

The fact that simultaneity at the same spatial point might no longer
be observer-independent exposes a possibly important limitation of the applicability
of our classical-spacetime analysis to a non-classical spacetime:
it was in fact crucial for our line of reasoning that we assumed that for
both observer $\mathcal{O}$ and observer $\mathcal{O}'$ the
emission of the two photons by the (point-like) source was simultaneous.
While this does not weaken our conclusion that only quantum-spacetime formulations
of DSR are admissible, it might weaken the insight on properties of these quantum spacetimes
that our analysis also provides.
Naively one might think that the needed quantum spacetimes should necessarily match
our result (\ref{DSRtransLawDeltat2}), but spacetime classicality was assumed in deriving
(\ref{DSRtransLawDeltat2}). Therefore it is too restrictive to insist precisely
on (\ref{DSRtransLawDeltat2}), if the analysis we here reported is ever generalized
to the case of (some example of) quantum spacetime.
Actually we feel that this is the direction where our analysis provides the most serious
challenge (and perhaps even an opportunity) for DSR research:
while it is true that the paradox we highlighted is not necessarily applicable
to quantum-spacetime scenarios, it is clearly at this point necessary
to address this issue within some specific quantum-spacetime picture and
in explicit physical fashion. Some explicit
candidate quantum-spacetime pictures for DSR with
energy-dependent speed of massless particles
have been proposed~\cite{gacmajid,kowaNCSTdisp,rainbowDSR},
but their analysis has not gone much further than the merely abstract/mathematical level,
whereas we here showed that a serious challenge (possibly excluding some or even all
of these candidates) will be met only upon a fully physical formulation
of these concepts of spacetime quantization.

\section{Closing remarks} \label{ClosingSec}
We have analyzed in this work two scenarios of departures from Lorentz invariance. Demanding the consistency of observations in different reference frames, we arrived at a characterization of the requirements from each scenario. Both scenarios require a costly departure from the present formulation of the laws of physics. For LSB we have the emergence of a ``privileged rest frame" in which $E_{QG}$ is defined and measured (and in all other frames we find that $E_{QG}$ is given by a specific non-trivial transformation). For DSR we find no classical solution to the consistency problem, as we have shown that the required transformation laws between boosted frames do not yield the Lorentz transformations as a limiting case. A potential resolution of the paradox is the departure from a classical spacetime and the possible emergence of apparently paradoxical spacetime nonclassicality.
The current literature on quantum-gravity finds advocates for each of these new-physics descriptions. Unfortunately, experiments of this nature, that will tell us which if any of these two possibilities should be adopted, cannot be imagined at present. However, we recall the recent history of the quantum-gravity research, when, just fifteen years ago, it was not imagined that experiments would at all be able to test Lorentz invariance violation at the Planck scale. \par

Still, even though an experimental test is unlikely at the immediate future, it is noteworthy what one finds by comparing the results (\ref{DeltaT2PrimoBroken}) and (\ref{DeltaT2PrimoPreserved}),
describing the behavior of the in-vacuo-dispersion-induced time delay $\Delta t_{II}'$ measured by observer $\mathcal{O'}$ w.r.t. the delay $\Delta t_I$ measured by $\mathcal{O}$,
as the relative speed between the two observers changes:
\begin{eqnarray} \label{PhenomComparison}
&{\Delta t_{II}' \simeq \mathcal{D}^{-1} \Delta t_{I}} \qquad ~~~~
&{\Delta t_{II}' \simeq \mathcal{D}^{n+1} \Delta t_I} . \\
&\text{(broken symmetry)} \qquad &\text{(deformed symmetry)} \nonumber
\end{eqnarray}
In this comparison we want to stress the relevant fact that in the two cases $\Delta t_{II}'$ behaves in the opposite way: when in the LSB case it shrinks w.r.t. $\Delta t_I$, in the DSR case it grows, and vice-versa.
If we consider again the conservative quantitative example of Sec. \ref{LSBsec}, we notice that even with present-day technology a possible measurement by two detectors would give a difference at a level of $\sim 0.1\%$ between the two scenarios in (\ref{PhenomComparison}). \par

While the result of the DSR analysis was derived in a classical-spacetime environment, which was shown to be an unfit description, it is probably safe to assume that the correct result, which may be derived when we have the full quantum-spacetime description, will admit
our result as at least a rough approximation.
Though we do not claim here that our DSR-case formula
should be trusted in detail, it appears likely that our preliminary analysis
has uncovered a qualitative feature that could be robust -
it might well be a valid assumption that
for observables such as $\Delta t_{II}'$ the dependence on the boost factor $\mathcal{D}$
of the DSR case would differ significantly
from the corresponding feature of the LSB case.
And while we have concentrated
here on a conceptual analysis,
it is not unreasonable to imagine
that if evidence of in-vacuo dispersion is ever found,
a two-telescope experiment could be conducted to investigate
further the departure from Lorentz invariance,
following the strategy of analysis we advocated here.
Since the contrast between the expected results of
the different modified-dispersion scenarios appears
to be a zeroth order effect, when the sensitivity
of time-of-flight tests reaches scales above the
quantum-gravity scale, such an experiment can in
principle be performed. This would help establish in detail the fate of Lorentz symmetry.

\begin{acknowledgments}
U.~J. thanks The University of Rome ``La Sapienza" for hospitality while some of this work was done.
G.~A.-C. is supported in part by grant RFP2-08-02 from The Foundational Questions Institute (fqxi.org).
T.~P. and U.~J. are supported in part by an ERC advanced research grant and by the Center of Excellence in High Energy Astrophysics of the Israel Science Foundation. U.~J. is also supported by the Lev-Zion fellowship of the Israel Council for Higher Education.
We are grateful to Lee Smolin for encouraging feedback on a draft of this manuscript and for forwarding to us a draft of the manuscript in \cite{leeTWOTIMES}.
\end{acknowledgments}

{}

\end{document}